# A scheme for total quantum teleportation

E. DelRe[1,3]*, B. Crosignani[2,3], and P. Di Porto[2,3]

[1]*Fondazione Ugo Bordoni, Via B.Castiglione 59, 00142 Roma, Italy*
[2]*Dipartimento di Fisica, Università dell'Aquila, L'Aquila, Italy*
[3]*Istituto Nazionale Fisica della Materia, Unità di Roma 1, Roma, Italy*

\* edelre@fub.it

**Abstract:** We address the issue of totally teleporting the quantum state of an external particle, as opposed to studies on partial teleportation of external single-particle states, total teleportation of coherent states and encoded single-particle states, and intramolecular teleportation of nuclear spin states. We find a set of commuting observables whose measurement directly projects onto the Bell-basis and discuss a possible experiment, based on two-photon absorption, allowing, for the first time, total teleportation of the state of a single external photon through a direct projective measurement.

Teleportation, in its present formulation, is an event by which a more or less complex physical object is transferred from one point in space to another without its actual material transportation. From a classical perspective, there is *no* conceptual impediment to teleportation: a system can be thoroughly scanned in a given location, and completely reconstructed in a second site, transmitting only the characterizing information but not the actual constituents. This procedure is evidently invalid for any *real* physical system, since its intrinsic quantum content cannot be generally measured and reconstructed. Remarkably, in the extreme (but *intrinsic* to any complex system) microscopic case of an unknown quantum state of a single external particle, Bennett et al. [1] have unveiled how quantum theory permits the "teleportation" of the state through the direct transmission of only classical information, whereas the quantum content is sent through an Einstein-Podolski-Rosen (EPR) two-particle entangled state [2]. In this formulation, two spin 1/2 particles 1 and 2 are prepared in the EPR singlet-state

$$\left|\Psi^-_{2,3}\right\rangle = \sqrt{1/2}\left(\left|\uparrow_2\right\rangle\left|\downarrow_3\right\rangle - \left|\downarrow_2\right\rangle\left|\uparrow_3\right\rangle\right). \tag{1}$$

Particle 2 is directed towards a region in which an experimenter called Alice is located, whereas the correlated particle 3 is sent to Bob. Alice is given a second (quantum) particle, labeled particle 1, wholly uncorrelated to 2 and 3, in an unknown spin state

$$\left|\phi_1\right\rangle = a\left|\uparrow_1\right\rangle + b\left|\downarrow_1\right\rangle, \tag{2}$$

where a and b are two unknown c numbers ($|a|^2 + |b|^2 = 1$). Let Alice perform an experiment on the system of particles 1 and 2 she has available, projecting on the Bell-operator basis [3] described by the two particle states

$$\left|\Psi^\pm_{1,2}\right\rangle = \sqrt{1/2}\left(\left|\uparrow_1\right\rangle\left|\downarrow_2\right\rangle \pm \left|\downarrow_1\right\rangle\left|\uparrow_2\right\rangle\right)$$
$$\left|\Phi^\pm_{1,2}\right\rangle = \sqrt{1/2}\left(\left|\uparrow_1\right\rangle\left|\uparrow_2\right\rangle \pm \left|\downarrow_1\right\rangle\left|\downarrow_2\right\rangle\right). \tag{3}$$

Since the three particle system is described by the wavefunction

$$\left|\Xi_{123}\right\rangle = \left|\phi_1\right\rangle\left|\Psi^-_{2,3}\right\rangle = (1/2)[\left|\Psi^-_{1,2}\right\rangle\left(-a\left|\uparrow_3\right\rangle - b\left|\downarrow_3\right\rangle\right) + \left|\Psi^+_{1,2}\right\rangle\left(-a\left|\uparrow_3\right\rangle + b\left|\downarrow_3\right\rangle\right)$$
$$+ \left|\Phi^-_{1,2}\right\rangle\left(b\left|\uparrow_3\right\rangle + a\left|\downarrow_3\right\rangle\right) + \left|\Phi^+_{1,2}\right\rangle\left(-b\left|\uparrow_3\right\rangle + a\left|\downarrow_3\right\rangle\right)], \tag{4}$$

particle 3 at Bob's end will be projected into a one-particle pure state corresponding to the outcome of Alice's measurement, that can be made identical to the initial unknown state $\left|\phi\right\rangle$ via a given unitary operation on the particle spin, after Alice has performed her measurement and has communicated her outcome to Bob, wherever he be situated.

Let us now consider the meaning of "external quantum state". Imagine the practically relevant case of Alice being asked to teleport a given physical object, for example the state (eq.(2)) of a *single* electron, about which she has no information. Any experiment directly performed by Alice on this system would give a *null* amount of information on the values of "a" and "b". For example, if she performs a spin measurement, and obtains $|\uparrow_1\rangle$, thus destroying $|\phi_1\rangle$ of which she has no replicas, she can only infer that the amplitude coefficient "a" is nonvanishing. Clearly, had Alice *prepared* the state $|\phi_1\rangle$, or, equivalently, had she replicas of it, the knowledge she could acquire on "a" and "b" would be greater, if not complete. In this paper, we consider the first, fundamental, case, i.e. a single particle whose origin is unknown to Alice. Teleportation in various conditions has been experimentally investigated in refs.[4,5,6,7], and will be discussed below.

Obtaining total teleportation of an unknown quantum state of a single particle requires that Alice carries out a set of measurements so as to project the state $|\Xi_{123}\rangle$ onto the basis of eq.(3). This task, far from being trivial, requires a careful analysis of necessary procedures, as done in refs.[8-9], where it is shown that, for the general case of a single particle, total teleportation cannot be achieved by means of linear operations. A possible nonlinear approach hinges on a two step indirect protocol, in the frame of nonlocal quantum measurements, where the initial Bell-states are disentangled into product states, which are successively measured [8-9] (see the discussion below of the experiment contained in ref.[6]).

Conversely, in this paper we address, for the first time, the Bell-measurements in a straightforward manner, through a direct application of basic principles of quantum mechanics. This necessarily implies the following steps:

(a) Determining a complete set of commuting observables possessing as common eigenvectors the four mutually orthogonal states given in eq.(3);

(b) Conceiving and realizing an experiment whose output corresponds to a common measurement of the above observables, i.e., to projecting $|\Xi_{123}\rangle$ onto one of the four terms on the R.H.S. of eq.(4).

Our approach, therefore, will be first to identify the relevant observables, and then to propose and discuss a possible related experiment.

In order to address point (a), we note that, while the components of the two-particle spin $\mathbf{S}_x = \mathbf{S}_{1x} + \mathbf{S}_{2x}$, $\mathbf{S}_y = \mathbf{S}_{1y} + \mathbf{S}_{2y}$, $\mathbf{S}_z = \mathbf{S}_{1z} + \mathbf{S}_{2z}$ do obviously not commute, this is not the case for $\mathbf{S}_x^2$, $\mathbf{S}_y^2$, $\mathbf{S}_z^2$. More precisely, $\mathbf{S}_x^2$, $\mathbf{S}_y^2$, $\mathbf{S}_z^2$ and $\mathbf{S}^2 = \mathbf{S}_x^2 + \mathbf{S}_y^2 + \mathbf{S}_z^2$ (scalar square of the total spin) are commuting observables (as can be easily deduced from elementary spin algebra) and the four states of eq.(3) are seen to be their eigenstates. The corresponding eigenvalue spectrum reads

$$
\begin{array}{c|cccc}
 & S^2 & S_x^2 & S_y^2 & S_z^2 \\
|\Psi^+\rangle & 2\hbar^2 & \hbar^2 & \hbar^2 & 0 \\
|\Psi^-\rangle & 0 & 0 & 0 & 0 \\
|\Phi^+\rangle & 2\hbar^2 & \hbar^2 & 0 & \hbar^2 \\
|\Phi^-\rangle & 2\hbar^2 & 0 & \hbar^2 & \hbar^2
\end{array}. \tag{5}
$$

By inspecting eq.(5), we note that the simplest minimal complete sets of commuting observables satisfying point (a) are the pairs $(\mathbf{S}_x^2, \mathbf{S}_y^2)$, $(\mathbf{S}_y^2, \mathbf{S}_z^2)$, and $(\mathbf{S}_z^2, \mathbf{S}_x^2)$. To be more explicit, if we choose to measure $\mathbf{S}_z^2$, $\mathbf{S}_x^2$, the output $S_z^2=0$, $S_x^2=\hbar^2$ corresponds to projecting particles 1 and 2 in $|\Xi_{123}\rangle$ onto $|\Psi^+_{1,2}\rangle$, $S_z^2=0$, $S_x^2=0$ corresponds to $|\Psi^-_{1,2}\rangle$, $S_z^2=\hbar^2$, $S_x^2=\hbar^2$ to $|\Phi^+_{1,2}\rangle$, and $S_z^2=\hbar^2$, $S_x^2=0$ to $|\Phi^-_{1,2}\rangle$, no other outputs being possible.

Let us now consider point (b): how do we perform an experiment aimed at measuring one of the above pairs? The most feasible apparata in the frame of teleportation are realized with photonic schemes [4,5,7] (see discussion in ref.[10]), although elaborate but promising schemes for atomic states have been proposed [11,12], and a short-range teleportation event has been observed via nuclear magnetic resonance [6]. A comparison of our proposal with experiments [4-7] will be discussed at the end of the paper. Hereafter, we propose a nonlinear optical protocol based on three two-photon absorption events [13] in cascade, as illustrated in Fig.1, and concentrate on one particular example based on atomic hydrogen spectroscopy, although other more or less elaborate realizations might be envisaged, possibly tailoring the transitions in artificial atoms [14]. Clearly, table of eq.(5) is not formally applicable to two-photon states. However, as shown below, the cascade measurement of the total angular momentum and of the square $S_z^2$ of its projection along the propagation axis, associated with an appropriate rotation, allows us to perform, in a direct analogy to point (a) and eq.(5), the desired direct projection operation on the four Bell states.

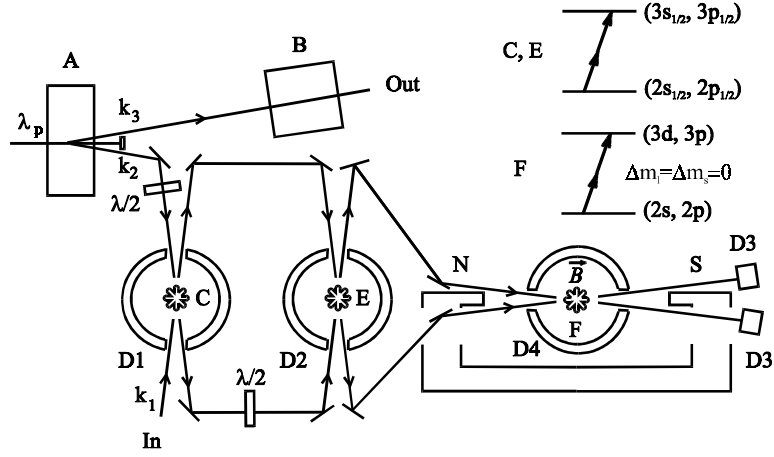

**Fig.1:** Total optical teleportation scheme: (A) PDC event; (B) linear polarization rotator; (C,E) singlet-selecting two-photon absorptions; (F) two-photon $\Delta m_s = \Delta m_l = 0$ absorption event (Zeeman configuration); (D1-D4) single photon detectors.

A two-photon polarization entangled state is generated by the parametric down-conversion (PDC) of a pump beam of wavelength $\lambda_p$ in a type II degenerate process in an appropriate nonlinear crystal (e.g., in a sample of β-Barium Borate). The pump $\lambda_p$ is chosen so as to correspond to the degenerate absorption lines $2s_{1/2}\{j=1/2, l=0\} \rightarrow 3s_{1/2}\{j=1/2, l=0\}$ and $2p_{1/2}\{j=1/2, l=1\} \rightarrow 3p_{1/2}\{j=1/2, l=1\}$ (taking into account spin-orbit coupling) of atomic hydrogen (i.e., $\lambda_p \approx 656.5$nm) [15]. The state generated after such an event in A can be described by [7]

$$|\gamma_{2,3}^-\rangle = \sqrt{1/2}\left(|R\rangle_{k_2}|R\rangle_{k_3} - |L\rangle_{k_2}|L\rangle_{k_3}\right) \tag{6}$$

(analogous to $|\Phi_{2,3}^-\rangle$) where $k_2$ and $k_3$ are the polarization entangled modes and R and L indicate respectively right and left-handed circular polarization, and the single photons have a $\lambda_s = 2\lambda_p$. In path $k_2$ we insert a $\lambda/2$ waveplate that rotates the polarization of the fields present in this mode according to the transformation [16]

$$|R\rangle_{k_2} \rightarrow |L\rangle_{k_2} \quad ; \quad |L\rangle_{k_2} \rightarrow -|R\rangle_{k_2} \tag{7}$$

so that the state of photons 2 and 3 is described by

$$|\chi_{2,3}^+\rangle = \sqrt{1/2}\left(|R\rangle_{k_2}|L\rangle_{k_3} + |L\rangle_{k_2}|R\rangle_{k_3}\right) , \tag{8}$$

analogous to $|\Psi_{2,3}^+\rangle$.

Photon 1, also of wavelength $\lambda_s$, is in an indeterminate and unknown single photon polarization state

$$|\phi\rangle_1 = a|R\rangle_{k_1} + b|L\rangle_{k_1} , \tag{9}$$

representing the state we wish to teleport [17].
The three mode system is described by the wavefunction

$$|\phi\rangle_1|\chi_{2,3}^+\rangle = (1/2)[|\chi_{1,2}^+\rangle\left(-a|L\rangle_{k_3} + b|R\rangle_{k_3}\right) + |\chi_{1,2}^-\rangle\left(-a|L\rangle_{k_3} - b|R\rangle_{k_3}\right) +$$
$$+ |\gamma_{1,2}^+\rangle\left(a|R\rangle_{k_3} - b|L\rangle_{k_3}\right) + |\gamma_{1,2}^-\rangle\left(a|R\rangle_{k_3} + b|L\rangle_{k_3}\right)] \tag{10}$$

where

$$\left|\chi^{\pm}_{1,2}\right\rangle = \sqrt{1/2}\left(\left|R\right\rangle_{k_1}\left|L\right\rangle_{k_2} \pm \left|L\right\rangle_{k_1}\left|R\right\rangle_{k_2}\right) \;\; ; \;\; \left|\gamma^{\pm}_{1,2}\right\rangle = \sqrt{1/2}\left(\left|R\right\rangle_{k_1}\left|R\right\rangle_{k_2} \pm \left|L\right\rangle_{k_1}\left|L\right\rangle_{k_2}\right). \qquad (11)$$

Modes $k_1$ and $k_2$ are made to approximately counterpropagate and cross in a region C where atomic hydrogen is concentrated. The region is subject to an external excitation, for example in the form of an ultraviolet lamp and secondary nonradiative processes or, alternatively, an electric discharge, promoting a relevant percentage of H atoms in the excited n=2 states. Given the value of $\lambda_s$ of the impinging photons, the H atoms are such as to allow for two-photon absorption from the $2s_{1/2}\{j=1/2, l=0\}$ to the $3s_{1/2}\{j=1/2, l=0\}$ and from the $2p_{1/2}\{j=1/2, l=1\}$ to the $3p_{1/2}\{j=1/2, l=1\}$, whereas all other transitions are nonresonant or not allowed, due to parity conservation. Note that counterpropagation is expedient to avoid Doppler-broadening [18], which could hamper wavelength selectivity of spin-orbit splitting, necessary in this part of the experiment. Furthermore, total angular momentum conservation implies that such transitions (implying the destruction of two photons) will occur only if the two-photon system is in a zero-spin state, that is if photons 1 and 2 are in state $\left|\chi^{-}_{1,2}\right\rangle$. Assuming a unitary process efficiency, if we place an ideal detector around the interaction region so as to collect the fluorescence (either primary or secondary) of the relaxation process, a detection is a signature that the two-photon system has been projected (on detector D1) on state $\left|\chi^{-}_{1,2}\right\rangle$ [19]. If this does not occur, modes $k_1$ and $k_2$ are still independent. In path $k_2$ we insert a second $\lambda/2$ waveplate leading to the transformation (see eqs.(7) and (11))

$$\left|\chi^{\pm}_{1,2}\right\rangle \rightarrow -\left|\gamma^{\mp}_{1,2}\right\rangle \;\; ; \;\; \left|\gamma^{\pm}_{1,2}\right\rangle \rightarrow \left|\chi^{\mp}_{1,2}\right\rangle \qquad (12)$$

The remaining system wavefunction is described by (apart from a normalization factor)

$$\left|\gamma^{-}_{1,2}\right\rangle\left(a\left|L\right\rangle_{k_3} - b\left|R\right\rangle_{k_3}\right) + \left|\chi^{-}_{1,2}\right\rangle\left(a\left|R\right\rangle_{k_3} - b\left|L\right\rangle_{k_3}\right) + \left|\chi^{+}_{1,2}\right\rangle\left(a\left|R\right\rangle_{k_3} + b\left|L\right\rangle_{k_3}\right). \qquad (13)$$

Both modes are now redirected onto a second region E, and again state $\left|\chi^{-}_{1,2}\right\rangle$ is selected by detector D2 in a process identical to process C. With respect to the initial wavefunction of eq.(10), this detector selects wavefunction $\left|\gamma^{+}_{1,2}\right\rangle$. The remaining system function is now

$$\left|\gamma^{-}_{1,2}\right\rangle\left(a\left|L\right\rangle_{k_3} - b\left|R\right\rangle_{k_3}\right) + \left|\chi^{+}_{1,2}\right\rangle\left(a\left|R\right\rangle_{k_3} + b\left|L\right\rangle_{k_3}\right). \qquad (14)$$

At this point, no further polarization rotation allows another identical two-photon process. The two remaining Bell states both transfer, in a two-photon absorption, two quanta of angular momentum ($S=2\hbar$). Furthermore, the two states are differentiated by the *square* $S_z^2$ of the total projection of the photon "spin" along the direction of propagation. In order to allow an $S_z^2$ measurement, we again concentrate, in region F, H atoms, in a configuration wholly identical to the two previous steps, but introduce a static magnetic field $B$ along the propagation z-axis, inducing Zeeman line splitting [15]. Furthermore, we make the two modes copropagate (as indicated in Fig.1) so as to activate Doppler broadening. By an appropriate choice of temperature we can energetically tune two-photon transitions from *all* the n=2 states to the n=3 ones, thus smearing the influence of **L**•**S** coupling (achieved via a Doppler broadening of approximately $\delta\nu_{Doppler} \approx 0.5 cm^{-1}$ for hydrogen n=2→n=3 levels [15]), as opposed to the previous part of the experiment. Since total spin conservation forbids the 2s-3s transition for the remaining $S=2\hbar$ photon-spin states, and parity conservation forbids the two-photon transitions 2s-3p and 2p-3s/3d, only a 2s-3d or a 2p-3p absorption can occur. For a strong enough Zeeman splitting, that is $\mu_B B > \delta\nu_{Doppler}$ (where $\mu_B$ is the Bohr magneton), corresponding to $B \approx 1T$, the line shifts are given by $\mu_B B(m_l+2m_s)$, where $m_l$ ($m_s$) is the orbital angular momentum (spin) projection along z. Therefore, energy tuning is preserved only for those transitions for which $m_l+2m_s$ is conserved. This implies that either both $m_l$ and $m_s$ are separately conserved or that $\Delta m_l = \pm 2$, $\Delta m_s = \mp 1$. These latter cases entail a global change of the total momentum projection on the z axis $\pm \hbar$, which cannot obviously be obtained through a two-photon absorption. Hence a two-photon transition, and thus an event on detector D4, must conserve both $m_l$ and $m_s$, so that it can take place only when the absorbed two-photon state is $\left|\chi_{1,2}^{+}\right\rangle$ ($\left|\gamma_{1,2}^{-}\right\rangle$ in the original eq.(10)), for which $S_z=0$ (angular momentum conservation). Detection on *both* detectors D3 implies finally that the state has been projected on the remaining state $\left|\gamma_{1,2}^{-}\right\rangle$ ($\left|\chi_{1,2}^{+}\right\rangle$ in the original eq.(10)), whereas detection on only *one* detector signals a noncoincidence. This last event indicates that either the PDC pair was available but the input state was not (detection on the "lower" D3), or that the input state was available, but the PDC pair was not (detection on the "top" D3).

Proceeding by way of the above processes, we are able to identify the linear polarization operation to be performed in B in order to obtain the initial polarization state on mode $k_3$ [1], realizing a total quantum teleportation event. The cascade events that involve two-photon nonlinear processes project on the basis of eq.(11), and thus constitute, in the photonic analogy, the operation of point (b). In other words, our measurements do not determine the sign of the two-particle spin projection, but only its absolute value, so that our procedure actually corresponds to determining $S_z^2$, as it must be. Had we measured the sign of $S_z$, we would have irreversibly reduced the conceptual maximum efficiency of the process, as in ref.[7]. Note that, since our treatment concerns a direct realization of the original teleportation scheme of ref.[1], it will work not only with pure states, but also with entangled EPR states. More precisely, Alice's original particle 1 can be itself part of an EPR entangled state along with another particle [1]. Clearly, in this case the state to be teleported is truly "unknown in principle".

Regarding the feasibility of this proposal, we note that the difficulties lie (i) in the generally small two-photon absorption probability, (ii) in the limited photon detection efficiencies. Concerning point (i), given an ordinary photon distribution, the probability of an H two-photon transition is extremely small [13][20]. However, we are concerned with those events in which the two photons *are* within a given pulse (the protocol in ref.[1]). If the pulse duration $\Delta\tau$ is comparable with the critical time $\tau_c$ of the absorption process, the probability of an event is much higher, and for feasible species concentrations and overlapping regions, a high transition probability is accessible. More precisely, the low ordinary two-photon absorption probabilities are essentially enhanced to typical single photon probabilities ($\sigma_{1S-2S,H} \approx 10^{-12} cm^2$, for the above mentioned atomic hydrogen two-photon transition) [21], thus allowing accessibly small absorption dimensions at standard conditions. Regarding point (ii), present commercial photon detection technologies allow *typical* single photon detection probabilities in excess of 70% (in the visible spectrum) [22].

Let us relate our scheme to previous experiments [4-7][23]. Refs.[4,6] address the problem of teleporting the quantum state of a single particle. In these experiments, the Authors do not measure $S_x^2$ and $S_z^2$ of which the Bell-states form the natural basis, but perform preliminary operations that project the two-particle 1 and 2 states into the disentangled basis (sometimes referred to as computational basis) [24] composed by the eigenstates of $S_{1z}$ and $S_{2z}$,

$$\left|\sigma_{1,2}^{++}\right\rangle=\left|\uparrow_1\right\rangle\left|\uparrow_2\right\rangle, \left|\sigma_{1,2}^{--}\right\rangle=\left|\downarrow_1\right\rangle\left|\downarrow_2\right\rangle, \left|\sigma_{1,2}^{+-}\right\rangle=\left|\uparrow_1\right\rangle\left|\downarrow_2\right\rangle, \left|\sigma_{1,2}^{-+}\right\rangle=\left|\downarrow_1\right\rangle\left|\downarrow_2\right\rangle, \qquad (15)$$

and then perform single particle measurements. Thus, the final projection on the Bell basis, realizing teleportation, requires a one-to-one correspondence between the two basis of eqs.(3) and (15). In ref.[4], this operation, which implies a passage from an entangled to a disentangled basis, is achieved simply because "particles" 1 and 2 actually refer to different degrees of freedom of the *same photon*, making both basis disentangled *a priori*. This is obtained at the expense of having to *encode* the state $\left|\phi\right\rangle$ we wish to teleport onto a local member of a two-photon momentum-entangled state, and thus this method, linear in nature, cannot be applied for the fundamental (and more general) case of teleporting the unknown (at least to Alice) state of an external particle, although it is conceptually apt to realize total teleportation of a *known* (prepared beforehand) state [8]. In ref.[6] this "rotation" (from basis of eq.(3) to eq.(15)) is possible through what is referred to as a "quantum exclusive-or gate" [24] or a "conditional spin flip" [8]. By this method, a nuclear spin state was teleported locally from an atom to another one belonging to the very same molecule. This transformation, that in principle allows total teleportation, *implies interaction among particles 1 and 2* (absent in photonic schemes), and, furthermore, implies conditional operations that *distinguish* between the two particles during their interaction. In our projective formulation these restraints are absent.

In ref.[5], a conceptually total unconditional (as opposed to conditional schemes [17]) teleportation of a coherent optical state is realized. In this experiment, one is able to reconstruct the coherent state given by a third party, Victor, to Alice, at Bob's station. Although this scheme succeeds in the relevant goal of teleporting the continuous quantum variables of a coherent state, it make use of inherently multi-particle states, so that it cannot be applied to tackle the issue of teleporting the discrete quantum state of a single particle [8].

In ref.[7], the teleported state truly refers to an external single particle state, in the sense of ref.[1]. Teleportation is achieved by means of a measurement of $S_z$, and thus is admittedly not total, and is obtained for at most 25% of all trials, and cannot possibly lead to a success rate higher than 50% (even though the actual efficiency of the process is lower). The conceptual limitation of this linear scheme is discussed in refs.[8-9], where it is stressed that the absence of "photon-photon" interaction prevents total teleportation of the unknown state of a single photon. In our formulation, this limitation, common to all linear processes, is an immediate consequence of the fact that Bell states do not form a basis of linear spin operators.

Our protocol, although not unrealistic as shown in its potential implementation with H atoms discussed above, presents relevant practical difficulties and is more complicated than previous experiments [4-7]. However, and this constitutes the fundamental import of this study, we stress that this experiment contains, in the most simple manner, all the ingredients necessary to obtain total teleportation of an unknown state of a single external particle, hinging on the measurements of the observables that project directly on the Bell basis (i.e. $S_x^2$, $S_z^2$). More precisely, the scheme presented in this paper is the first one that allows, in principle, *total teleportation of the external state of a single photon*.

**References:**

[1]   C.H.Bennett et al., *Phys.Rev.Lett.* **70**, 1895 (1993).
[2]   A.Einstein, B.Podolsky, and N.Rosen, *Phys.Rev.* **47**, 777 (1935).
[3]   S.L.Braunstein, A.Mann, and M.Revzen, *Phys.Rev.Lett.* **68**, 3259 (1992).
[4]   D.Boschi et al., *Phys.Rev.Lett.* **80**, 1121 (1998).
[5]   A.Furusawa et al., *Science* **282**, 706 (1998).
[6]   M.A. Nielsen, E.Knill, and R.Laflamme, *Nature* **396**, 52 (1998).
[7]   D.Bouwmeester et al., *Nature*, **390**, 575 (1997).
[8]   L.Vaidman and N.Yoran, *Phys.Rev.A* **59**, 116 (1999).
[9]   N.Lutkenhaus, J.Calsamiglia, and K.-A.Suominen, *Phys.Rev.A* **59**, 3295 (1999).
[10]  D.N.Klyshko, *Physics-Uspekhi* **41**, 885 (1998).
[11]  L.Davidovich et al. *Phys.Rev.A* **50**, R895 (1994).
[12]  J.I.Cirac, and A.S.Parkins, *Phys.Rev.A* **50**, R4441 (1994).
[13]  R.Loudon, *The Quantum Theory of Light* (Clarendon Press, Oxford, 1983).
[14]  regarding two-photon absorption, see, for example, the review by R.Cingolani and K.Ploog, *Adv.Phys.* **40**, 535 (1991).
[15]  see, for example, B.H.Bransden and C.J.Joachain, *Physics of Atoms and Molecules*, (Longman, Essex, 1983).
[16]  A.Yariv and P.Yeh, *Optical Waves in Crystals* (Wiley, New York, 1984)
[17]  "Single-photon" schemes based on parametric downconversion (PDC), are practically limited to so-called "coincidence experiments"[see refs. 4,7, and 10] in which *both* the two-photon state given in eq.(8) and the state of eq.(9) are *present* (in a given time interval). This handicap, based on the present limitations of reliable sources of entanglement, does not however constitute a conceptual limitation to the observation of a single total quantum teleportation event.
[18]  C.L.Cesar et al., *Phys.Rev.Lett.* **77**, 255 (1996).
[19]  Undesired detection events of photons due to spontaneous transitions of the type 2p-1s, or to photons emitted directly by the UV lamp, can be suppressed through appropriate wavelength filtering at the detectors associated with band-pass selection of lamp emission or discharge excitation.
[20]  P.W.H.Pinkse et al. *Phys.Rev.Lett.* **79**, 2423 (1997).
[21]  H.Fei et al. *Phys.Rev.Lett.* **78**, 1679 (1997).
[22]  APD SPCM-AQ (EG&G Optoelectronics), product data sheet (1999).
[23]  We highlight the theoretical limitations to teleporting schemes, and we do not refer to the actual limits of experiments, connected to measured fidelity.
[24]  G.Brassard, S.Braunstein, and R.Cleve, *Physica D* **120**, 43 (1998).